\documentclass[useAMS, usenatbib, fleqn]{mn2e}
\usepackage{graphicx}
\usepackage{epstopdf}
\usepackage{amsmath}

\def\aaps{Astronomy \& Astrophysics, Supplement}
\def\apss{Astrophys. Space. Sci.}
\def\apj{APJ}
\def\apjs{The Astrophysical Journal, Supplement}
\def\apjl{APJ, Part 2 - Letters}
\def\aap{ Ann. Phys. (Paris)}

\def\mnras{MNRAS}
\def\nat{Nature}

\title{Adiabatic Evolution of Mass-losing Stars}

\author[Lixin (Jane) Dai, Roger D. Blandford, \& P. P. Eggleton]{Lixin Dai$^{1}$\thanks{E-mail:
cosimo@stanford.edu (LD)}, Roger D. Blandford$^{2}$ \thanks{E-mail: rdb3@stanford.edu (RB)} and P. P. Eggleton$^{3}$\thanks{E-mail: eggleton1@llnl.gov (PE)} \\
$^{1,2}$Kavli Institute for Particle Astrophysics and Cosmology, Stanford University, Menlo Park, CA 94025, USA\\
$^{3}$Lawrence Livermore National Laboratory, 7000 East Ave, Livermore, CA 94551, USA\\}

\begin{document}

\pagerange{\pageref{firstpage}--\pageref{lastpage}} \pubyear{2009}

\label{firstpage}

\maketitle

\begin{abstract}
We have calculated the equilibrium properties of a star in a circular, equatorial orbit about a Super-Massive Black Hole (SMBH), when the star fills and overflows its Roche lobe. The mass transfer time scale is anticipated to be long compared with the dynamical time and short compared with the thermal time of the star, so that the entropy as a function of the interior mass is conserved. We have studied how the stellar entropy, pressure, radius, mean density, and orbital angular momentum vary when the star is  evolved adiabatically, for a representative set of stars. We have shown that the stellar orbits change with the stellar mean density. Therefore, sun-like stars, upper main sequence stars and red giants will spiral inward and then outward with respect to the hole in this stable mass transfer process, while lower main sequence stars, brown dwarfs and white dwarfs will always spiral outward. 

\end{abstract}

\begin{keywords}
stars: mass-loss, stars: evolution, stars: kinematics and dynamics, binaries, accretion, galaxies: active
\end{keywords}

\section{Introduction}

When a star loses mass on a time scale faster than the Kelvin-Helmholtz time (but 
slower than the dynamical time), as can happen in Extreme Mass-Ratio Inspirals (EMRI), 
the structure of the star will evolve adiabatically so that the entropy as a function 
of interior mass $S(m)$ is approximately conserved. This will be true of radiative and 
convective zones (except near the surface). \citet{Webbink85} gave a general 
introduction to binary mass transfer on various time scales. \citet{Hjellming87} used 
polytropic stellar models to explore the stability of this adiabatic process, and 
listed several critical mass ratios above which the donor stars are unstable on 
dynamical time scales. \citet{Soberman97} further discussed the stability of binary 
system mass transfer processes on the thermal and dynamical times scales.

The previous works generally assumed comparable masses of the two objects 
in the binary system, and an unchanging distance between them throughout the 
process. In this paper, we will use real stellar models to study how stars respond to the 
loss of mass adiabatically on a time scale slower than the dynamical time scale but 
still faster than the thermal one. The star's orbital radius from the hole is no 
longer held as a constant. The way that the star evolves in such a mass-transfer 
environment depends upon the mean density of the stripped star as a function of 
decreasing mass. Nuclear reactions will be shut off as soon as the mass loss starts, 
as they are highly temperature-sensitive \citep{Woosley} and the central temperature 
decreases, at least in the cases that concern us here. 

In our EMRI binary system, the star is assumed to be in a circular, equatorial orbit 
about the central massive black hole. The mass transfer starts when the star just 
fills its Roche lobe, when materials will flow out of the inner Lagrangian point L1. We call this the Roche mass transfer process. During the mass transfer phase, the stellar orbital radius may increase or decrease. However, we assume that the change is slow enough so that the stellar orbit remains circular. If the orbit crosses the black hole's innermost stable circular orbit (ISCO) before it is tidally consumed, it will plunge into the hole. The ISCO has radius 6$R_g$ for a non-rotating black hole, and 1.237$R_g$ for a spin parameter $a =0.998$ black hole \citep{Thorne:74}. Here $R_g $ is the gravitational radius of the hole defined as $GM_{\rm{BH}}/ c^2$, where $G$ is the gravitational constant, $M_{\rm{BH}}$ is the mass of the black hole, and $c$ is the speed of light.

In this paper we consider the evolution of a representative set of stars and planets 
under these conditions. In section 2 we discuss the radius-mass relation of a star 
composed of ideal gas when losing mass adiabatically. In section 3 we study the case 
of a solar type star. In section 4 we investigate lower main sequence stars, upper 
main sequence stars, red giants, white dwarfs, brown dwarfs and planets. We will 
illustrate the use of these evolutionary models in upcoming papers. 

After this work was completed, our attention was drawn to a recent paper: \citet{Ge}. 
Although the end application of their paper is different, some of the calculations 
overlap (and agree with) those presented below.

\section{Radius-Mass Relation}

\begin{figure}
 \centering
 \includegraphics[width=3in]{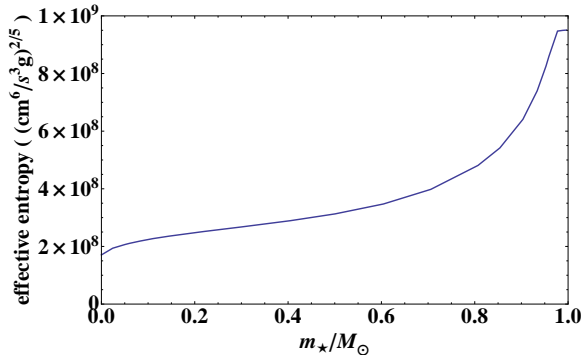}
 \caption{
The local entropy of the Sun as a function of its interior mass. The effective entropy 
is defined as $P^{3/5} \rho^{-1}$. The curve has a flat portion at the upper right, 
since the entropy is almost constant in the surface convective zone.} 
 \label{entropy}
\end{figure}

For a star composed of ideal gas, we have:
\begin{equation}
   \frac{dr}{dm} = \frac{1}{4 \pi r^2 \rho(m)},
\end{equation}
where $r$ is the radius, $m$ is the interior mass, and $\rho$ is the density. This 
gives
\begin{equation}
   \frac{dP}{dm} = - \frac{Gm}{4 \pi r^4},
\end{equation}
where $P$ is the gas pressure, and $G$ is the gravitational constant.

Ignoring radiative contributions, Coulomb and degeneracy corrections, the entropy per 
particle will be $S/N \sim C_v \ln \left(P/\rho^{\gamma} \right)$ with $\gamma = 5/3$ 
for ideal monatomic gas, up to a constant; and we can define an effective entropy 
$\tilde{S} = P^{3/5} \rho^{-1}$ for such an ideal gas.

Then, for a mass-losing star, with constant local entropy, we have
\begin{eqnarray}
\label{masslosing}
   \frac{dr}{dm} &=& \frac{\tilde{S} (m)}{4 \pi r^2 P^{\frac{3}{5}}}, \nonumber\\
   \frac{dP}{dm} &=& - \frac{Gm}{4 \pi r^4}.
\end{eqnarray}

For a star with a known entropy profile, we can solve for its new equation of state 
with boundary conditions that $r_\star(0)=0$ and $P_\star(0)=P_{0 \star}$. We solve 
for the total stellar mass $M_\star$ and the surface radius 
$R_\star = r_\star(M_\star)$, where $P_\star(M_\star)=0$. Then the volume and average 
density of the star in this new model can be obtained.

\begin{figure}
 \centering
 \includegraphics[width=3in]{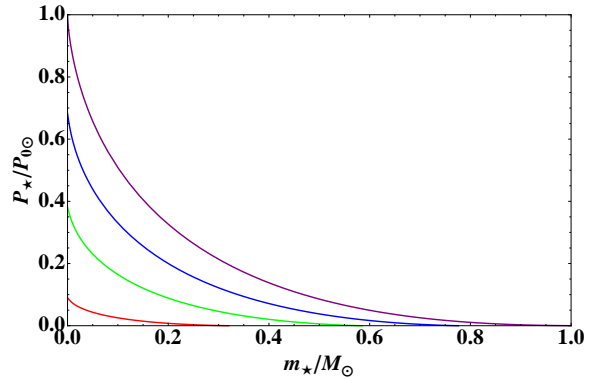}
 \caption{  The pressure vs interior mass for a mass-losing Sun-like star.  The four 
colored curves represent the four cases when the stellar central pressure decreases to 
$ 0.7 \  \rm{(blue)} , \ 0.4 \  \rm{(green)}, \  \rm{and} \  0.1 \  \rm{(red)} 
\ \rm{from} \ 1  \  \rm{(purple)} \ P_{0 \odot}$. } 
 \label{PMScaled}
\end{figure}

\begin{figure}
 \centering
 \includegraphics[width=3in]{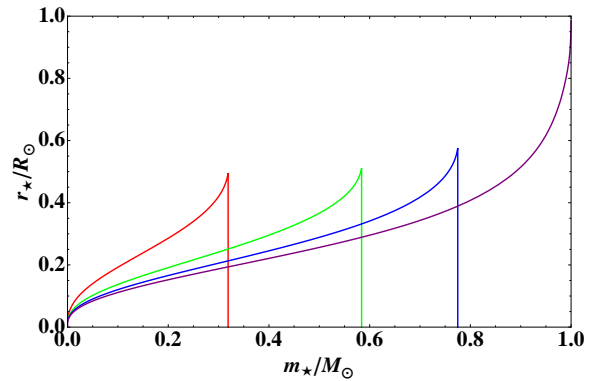}
 \caption{ The radius vs interior mass for a mass-losing Sun-like star. The four 
curves represent the same cases as in Fig. \ref{PMScaled}. } 
 \label{RMScaled}
\end{figure}

When the star orbits in a circular equatorial orbit and loses mass under stable 
evolution during this Roche mass transfer phase, its orbital period will follow 
$P_R \propto {\bar{\rho}_\star}^{-1/2}$, where $P_R$ is the Roche orbital period, and 
$\bar{\rho}_\star$ is the mean density of the stripped star. This formula can be 
obtained by comparing the tidal force from the hole and the gravitational force from 
the star for a point mass on the Roche surface, and also using Kepler's laws in the 
Newtonian limits. Through simple calculation we can also get the angular momentum of 
the system $L_\star \sim m_\star P_R^{1/3} \sim m_\star {\bar{\rho}_\star}^{-1/6}$, 
which should decrease for stability throughout the Roche mass transfer phase through 
slowly gravitational radiation. Here we neglected the much smaller angular momentum 
carried away by the hot stream flowing out of the inner Lagrangian point (L1). 
Therefore, the change of ${\bar{\rho}}_\star$ and $L_\star$  in this adiabatic 
evolution would allow us to know whether the mass-transfer is stable or catastrophic 
on the dynamical time scales, and also how the stellar orbit evolves through this 
process. For example, if the mean density of the star decreases, its orbital period 
will increase indicating that it is moving farther from the hole. The exact formulas 
of $P_R$ and $L_\star$ will be discussed in upcoming papers, in not only the Newtonian 
limits but also the relativistic limits.

\section{Solar Model}

\begin{figure}
 \centering
 \includegraphics[width=3in]{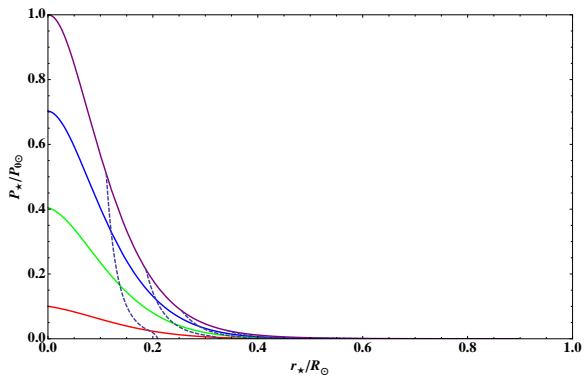}
 \caption{ The pressure vs radius for a mass-losing Sun-like star. The four solid 
curves still represent the cases when a Sun-like star loses its mass until its new 
central pressure is $P_{0 \star} = 1,  \ 0.7,  \ 0.4, $ and $0.1 \ P_{0 \odot}$, from the 
top (purple) to the bottom (red). The five blue dashed lines are contours enclosing 
the same interior mass: $0.1, \ 0.3,  \ 0.5,  \ 0.7,\ \rm{and} \ 0.9 \ M_\odot$ from 
left to right respectively. (The two rightmost lines are very hard to see). We can 
observe that as the star loses its mass, the surface pressure of any shell decreases 
and that shell enclosing constant mass expands.}
 \label{PRcontourScaled}
\end{figure}

\begin{figure}
 \centering
 \includegraphics[width=3in]{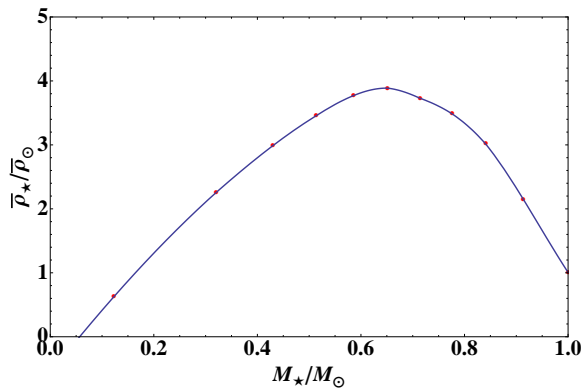}
 \caption{The mean density of a star with decreasing total mass for a Sun-like star. 
The blue solid curve is an interpolation of the red data points for $P_{0 \star } = 
0.01,\ 0.1, \ 0.2, \ 0.3, \ 0.4, $ $ 0.5, \ 0.6, \ 0.7, \ 0.8, \ 0.9 \ $ and 
$1 \ P_{0 \odot}$ from left to right.} 
 \label{density_sun}
\end{figure}

Let us start with the Sun as a representative example of main sequence stars. We 
adopted Guenther and Demarque's standard solar model (\citet{Demarque, Guenther}) to 
derive $\tilde{S}(m)$ as in Fig.\ref{entropy}. Notice that the entropy is almost 
constant for the outer thirty percent of its radius, which corresponds to the convective zone 
of the Sun.

Using the entropy profile, we computed the radius of the star $r_\star$ and the 
pressure of the star $P_\star$ as functions of the interior mass $m_\star$, when 
the star is adiabatically evolved to a new equilibrium.  We plotted  
$r_\star (m_\star)$, $P_\star (m_\star)$,and $P_\star(r_\star)$ in Fig.\ref{PMScaled}, 
Fig. \ref{RMScaled}, and Fig.\ref{PRcontourScaled}, as the star loses mass to the 
point that the central pressure equals $0.7, \ 0.4, \ 0.1$ from the original stellar 
central pressure $P_{0 \odot}$. As the mass of the star is stripped, its interior 
pressure decreases and total radius shrinks. 

As the mass of a Sun-like star decreases, its central pressure also decreases. By 
observing contour lines of the constant interior masses in Fig.\ref{PRcontourScaled}, 
we can see that shells enclosing the constant masses are expanding in this evolution.

The total stellar volume, however, is still almost monotonically decreasing in this 
evolution. For a solar model, the volume contracts first and then remains almost 
constant. Therefore its mean density increases first and then drops. We can observe 
the change of the stellar mean density as the mass decreases in Fig. 
\ref{density_sun}. For the first $\sim 40 \% $ of the solar mass, the star moves 
closer to the hole since the density is increasing, then the star moves farther from 
the hole for the rest of the Roche mass transfer phase.

\begin{figure}
 \centering
 \includegraphics[width=3in]{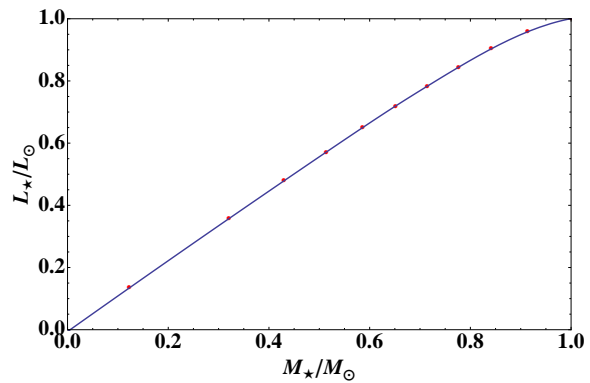}
 \caption{ The effective angular momentum $L_\star$ of the star with decreasing total 
mass for a Sun-like star.  $L_\star$ is defined as  $M_\star \bar{\rho}_\star ^{-1/6}$,
 and $L_\odot$ is defined as $M_\odot \bar{\rho}_\odot^{-1/6}$. The data points 
represent the same cases as in Fig. \ref{density_sun}.} 
 \label{L_sun}
\end{figure}

Lastly we want to check whether this Roche process is stable or not. By plotting the 
effective angular momentum $L_\star$ as a function of the rest total stellar mass 
$M_\star$ as in Fig. \ref{L_sun}, we confirmed that $L_\star$ does decrease. The mass transfer is stable on a dynamical time scale.

\section{Other Stars}

\subsection{Lower Main Sequence Stars}

Lower main sequence stars have different structures from the Sun. The whole star is 
convective and has almost constant entropy (e.g., \citet{Girardi00, Padmanabhan}). 
Here we chose a star with $\sim 0.3$ stellar mass as a representative. We adopted models computed by one os us (PE) for other purposes. All these zero age main 
sequence stars have metalicity $Z=0.01$.

\begin{figure}
 \centering
 \includegraphics[width=3in]{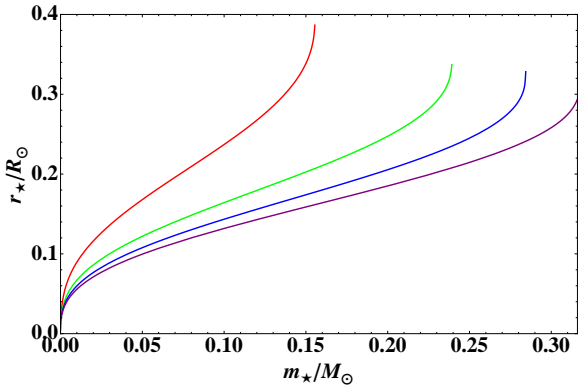}
 \caption{
The radius vs interior mass for a mass-losing lower main sequence star with 
$ \sim 0.3 M_\odot $ mass. The four colored curves represent the four cases when the 
stellar central pressure equals $0.1 \ \rm{(red)}, \ 0.4 \ \rm{(green)}, \ 0.7 \ \ 
\rm{(blue)}, \ \rm{and} \ 1  \ \rm{(purple)}$ original stellar central pressure. } 
 \label{RM0.316}
\end{figure}

\begin{figure}
 \centering
 \includegraphics[width=3in]{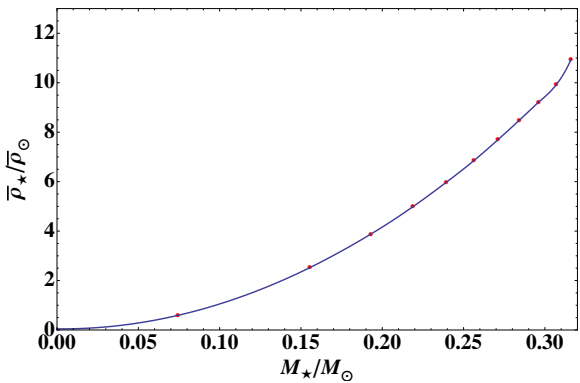}
 \caption{ The stellar mean density with decreasing total mass for a lower main 
sequence ($\sim 0.3 M_\odot$) star. The blue solid curve is an interpolation of the 
red data points representing $P_{0 \star } = 0.01,\ 0.1, \ 0.2, \ 0.3, \ 0.4,  \ 0.5, 
\ 0.6, \ 0.7, \ 0.8, \ 0.9$ and $1$ original stellar central pressure, from left to 
right. } 
 \label{Density0.316}
\end{figure}

As the star loses its mass, it expands and its interior pressure decreases, similar to 
the solar case. The radius-mass relation (Fig. \ref{RM0.316}) shows that the major 
difference between this model and the solar model is that the total volume of a 
low-mass star is always monotonically increasing. Therefore, its mean density will 
continue to decrease in the Roche phase (Fig. \ref{Density0.316}). A lower 
main sequence star will move out as it transfers mass. The effective angular 
momentum for this star however decreases throughout the Roche mass transfer phase due 
to the loss of mass, confirming that the star is stripped stably on the dynamical 
time scale.

\subsection{Upper Main Sequence Stars}
Upper main sequence stars have convective cores for more efficient energy transport 
(e.g., \citet{Woosley}). This mixing of material around the core removes the helium 
ash from the hydrogen burning region, allowing more of the hydrogen in the star to be 
consumed during the main sequence lifetime. Therefore the local effective entropy is 
almost constant in the center. The outer regions of a massive star are radiative. We 
investigated a $\sim 7.9 \ M_\odot$ star as an example.

\begin{figure}
 \centering
 \includegraphics[width=3in]{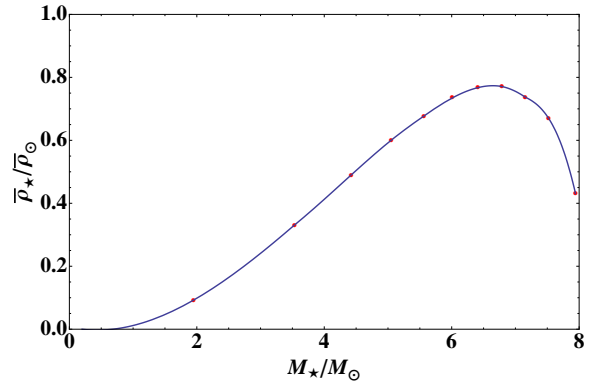}
 \caption{The stellar mean density with decreasing total mass for an upper main 
sequence star with $\sim 7.9 M_\odot$ mass). The blue solid curve is an interpolation 
of the red data points representing $P_{0 \star } = 0.01,\ 0.1, \ 0.2, \ 0.3, \ 0.4, 
\ 0.5, \ 0.6, \ 0.7, \ 0.8, \ 0.9$ and $1$ original stellar central pressure, from 
left to right.}
 \label{Density7.943}
\end{figure}

The mean density of an upper main sequence star behaves similarly to the solar case, 
as shown in Fig. \ref{Density7.943}. An upper main sequence star will first move in and 
then out, while stably filling its Roche lobe in the mass-accreting phase, since its 
angular momentum can be calculated as decreasing as well.

\subsection{Red Giants}
\begin{figure}
 \centering
 \includegraphics[width=3in]{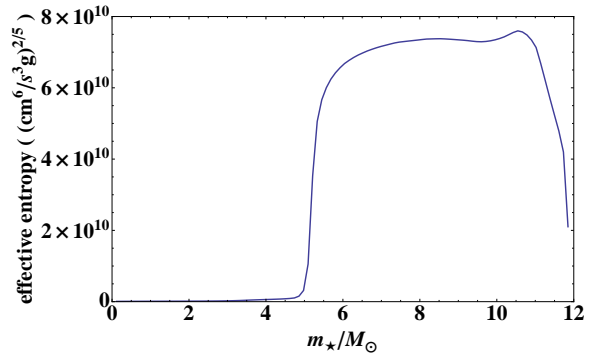}
 \caption{
The local entropy of a red supergiant of mass $11.85 \ M_\odot$ as a function of the 
stellar interior mass. A carbon core resides mostly within $\sim 2.7 \ M_\odot$, 
followed by a helium core lasting untill $\sim 6\ M_\odot$. The entropy for the 
hydrogen shell outside is much higher due to its low density.} 
 \label{entropyrg}
\end{figure}

Main sequence stars above $0.5 \ M_\odot$ evolve into red giants in their late phases 
\citep{Reimers, Sweigart}. Here we actually used a red supergiant with 
$11.85 \ M_\odot$ to study how it reacts to the loss of mass due to tidal stripping. 
Its effective entropy, defined in the same manner as before, is shown in Fig. 
\ref{entropyrg}. We can see a carbon-helium core with higher density and a much less 
dense hydrogen shell with $\sim 6 \ M_\odot$ mass inside.

\begin{figure}
 \centering
 \includegraphics[width=3in]{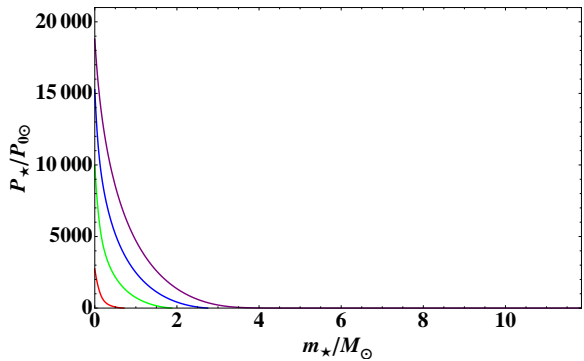}
 \caption{
The pressure vs interior mass for the red supergiant of Fig. \ref{entropyrg}. The four curves are cases when the 
stellar central pressure equals $0.1 \ \rm{(red)}, \ 0.4 \ \rm{(green)}, \ 0.7 \ \ 
\rm{(blue)}, \ \rm{and} \ 1  \ \rm{(purple)}$ of the original central pressure. The 
pressure decreases very fast going from the center of the star to its surface. 
There is a very extended hydrogen outer layer with relatively much lower density, as 
can be seen from the purple curve. } 
 \label{PMrg}
\end{figure}

\begin{figure}
 \centering
 \includegraphics[width=3in]{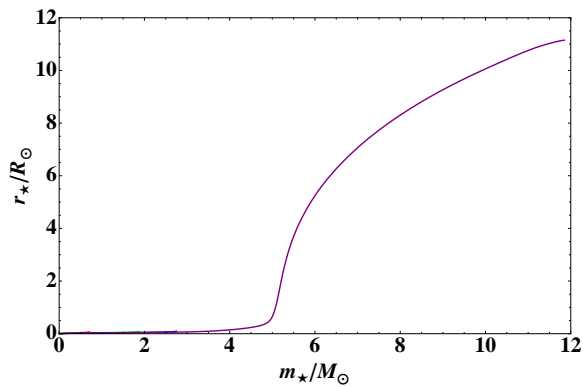}
 \caption{
The radius vs interior mass for the red supergiant of Fig. 10. The four cases correspond to the same 
models as in Fig. \ref{PMrg}. We can see that the volume of the red supergiant shrinks a lot soon 
after the tidal stripping starts. Also losing the hydrogen photosphere leaves the 
central region of the star almost unaffected.} 
 \label{RMrg}
\end{figure}

\begin{figure}
 \centering
 \includegraphics[width=3in]{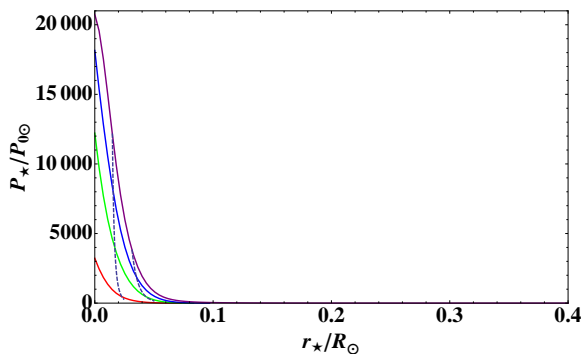}
 \caption{
The pressure vs radius profile of a mass-losing red supergiant with total initial 
mass $11.85 \ M_\odot$. Here the dashed lines represent the shell enclosing 
$0.02 \ M_\star$ (left) and $0.1 \ M_\star$ (right).}
 \label{PRcontourrg}
\end{figure}

The pressure - mass, radius - mass, and pressure - radius relations are also plotted 
in Fig.\ref{PMrg}, Fig.\ref{RMrg}, and Fig. \ref{PRcontourrg}. We can observe that 
red giants have pressure-mass curves with much steeper gradients compared with 
main sequence stars, and very extended low-density hydrogen outer layers. When this 
hydrogen shell is fed to the SMBH very rapidly, the central pressure is not affected 
much. In that regime, the response of the star is not strictly adiabatic.

\begin{figure}
 \centering
 \includegraphics[width=3in]{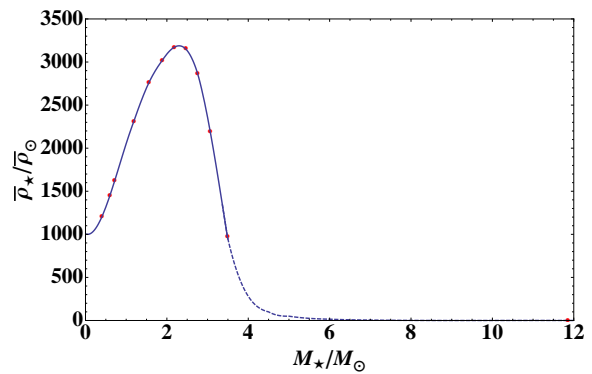}
 \caption{
The mean stellar density with decreasing total mass for a red supergiant with total initial mass $11.85 \ M_\odot$. The red data points stand for $P_{0 \star }$ equals  $0.05,\ 0.08, \ 0.1, \ 0.2, \ 0.3, \ 0.4, \ 0.5, \ 0.6, \ 0.7, \ 0.8, \ 0.9$ and $1$ original stellar central pressure. The mean densities for $M 3.5 M_\odot$ are probably inaccurate owing to the relative fragility of the red giant envelope.} 
 \label{MRrg}
\end{figure}

We also plotted Fig. \ref{MRrg} to see how the mean density of the red giant changes 
with its total mass. The star's total volume shrinks by a factor bigger than a 
thousand as the hydrogen envelope is consumed, therefore the mean density increases a 
lot at the beginning and then starts to decrease. The red giant is stably stripped, 
while first moving in and then out.

A star is a red giant for only a small fraction ($10\%$ to $25\%$) of its fusion 
lifetime (e.g., \citet{Padmanabhan}). Therefore, it would be rare to observe a red giant 
feeding an SMBH accretion disc.

\subsection{White Dwarfs}

White dwarfs are composed of a nondegenerate gas of ions and a degenerate and at least 
partly relativsitc gas of electrons \citep{Kippenhahn}. We need to redefine its 
effective entropy since the entropy of a degenerate electron gas differs from that of 
a nondegenerate gas. However, an easier way of obtaining the structure of the evolved 
white dwarf would be just to use its radius-mass relation:

\begin{equation}
   R_{\star} (M_{\star}) \simeq 0.022 \mu_e^{-1} R_\odot \left( \frac{M_\star}
{M_{\rm{ch}}} \right)^{- \frac{1}{3}} \left[ 1- \left( \frac{M_\star}{M_{\rm{ch}}} 
\right) ^{\frac{4}{3}} \right]^{\frac{1}{2}}
\end{equation}
\citep{Nauenberg72}.
In this equation, $R_\star$ and $M_\star$ are the total radius and mass of the evolved 
white dwarf. $\mu_e$ is the average number of nucleons per electron, and $\mu_e=2$ for 
He, C, O, which is also a good approximation for most other options. $M_{\rm{ch}}$ is 
the Chandrasekhar mass, which is approximately $1.459 M_\odot$. The formula holds 
true, except when the stellar mass is close to its upper limit $M_{\rm{ch}}$, where 
the radius should approach a constant.

\begin{figure}
 \centering
 \includegraphics[width=3in]{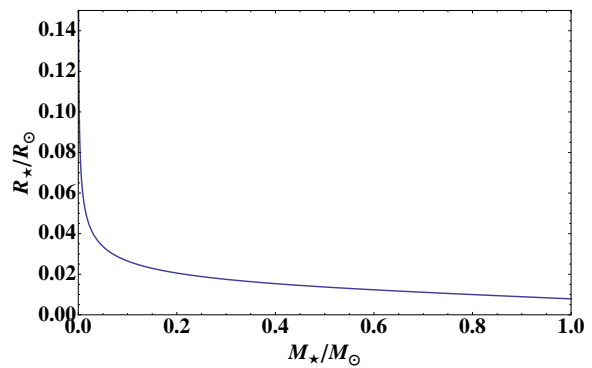}
 \caption{
The total radius as a function of the total mass for a white dwarf, as the total mass 
decreases from $\bf{1 M_\odot}$ to zero. We can see that the less massive a white 
dwarf is, the larger its volume actually is.} 
 \label{RM_wd}
\end{figure}

\begin{figure}
 \centering
 \includegraphics[width=3in]{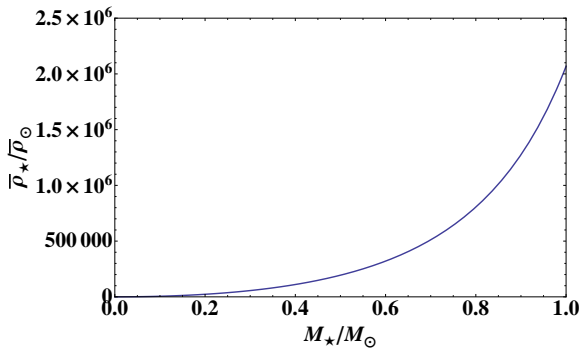}
 \caption{
The stellar mean density of a white dwarf, as its total mass  
decreases from $1M_\odot$ to zero. As its mass is stripped, its mean density decreases.} 
 \label{density_wd}
\end{figure}

We consider a $1 M_\odot$ white dwarf with complete degeneracy and zero temperature, 
and show its radius-mass and mean density-mass relations in Fig. \ref{RM_wd} and 
Fig.\ref{density_wd}. As the white dwarf is tidally stripped, its total volume 
increases in this adiabatic evolution, ensuring that the mean stellar density 
decreases throughout the Roche phase. Therefore, as a white dwarf loses its mass under 
the Roche process, it will always expand its orbit. 

For a white dwarf with a mass close to $M_{\rm{ch}}$, its radius does not change much 
as it loses mass, so the effective angular momentum ($L_\star \propto M_\star 
\bar{\rho_\star}^{-1/6} \propto M_\star ^{5/6}$) decreases throughout the Roche 
mass-transfer phase.  Adiabatic mass-transfer is stable on the dynamical time scale 
for all types of white dwarfs.

\subsection{Brown Dwarfs}
Brown dwarfs never undergo hydrogen reactions. They are supported by the degeneracy 
pressure of their non-relativistic electrons \citep{Kippenhahn}. Their ions are treated as ideal gases. Brown dwarfs are also fully 
convective.

The total pressure of a brown dwarf, including the ideal ion gas pressure and the 
electron degeneracy pressure, goes as:
 \begin{equation}
   P \simeq K \rho^{\frac{5}{3}},
\end{equation}
where $K$ is a constant close to $10^{13}$ dyn cm$^{-2}$ (e.g., \citet{Padmanabhan, 
Burrows93, Burrows01, Burrows02}; and the references therein). The effective entropy, 
still defined as $P^{3/5} \rho^{-1}$, therefore would be a constant, confirming the 
convective structure.

\begin{figure}
 \centering
 \includegraphics[width=3in]{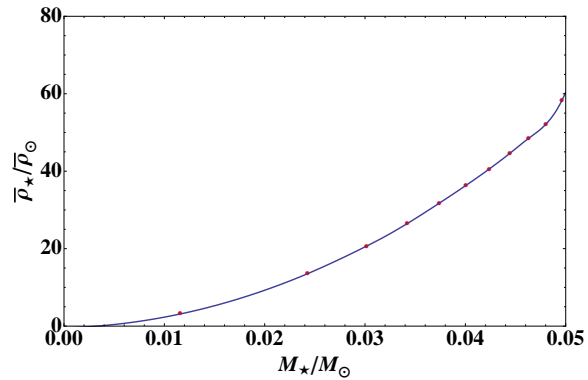}
 \caption{
 The stellar mean density as a function of its total mass for a brown dwarf, as its 
total mass decreases from $0.05 M_\odot$ to zero. We can see that the less massive a 
brown dwarf is, the less dense it is in this adiabatic evolution. The red data points 
were obtained using the model with  $P_{0 \star } = 0.01,\ 0.1, \ 0.2, \ 0.3, \ 0.4,  
\ 0.5, \ 0.6, \ 0.7, \ 0.8, \ 0.9$, \ and $1$ original stellar central pressure, from 
left to right, and the blue solid curve is an interpolation of these data points.} 
 \label{density_bd}
\end{figure}

Using the relation between $P$ and $\rho$, we can solve the equations of equilibrium 
[eq. (\ref{masslosing})] for a brown dwarf with a certain new central density and 
total mass. Here we used a 1 Gyr brown dwarf with total mass $0.05 M_\odot$ with 
central density $458.0$ g cm$^{-3}$ and total radius $6.4 \times 10^9$ cm \citep{Burrows93}. K is 
calculated to be $\sim 8.5 \times 10^{12}$ dyn cm$^{-2}$, in order to fit the model. 
Using our routine, we show how the mean stellar density varies with total mass 
in Fig. \ref{density_bd}. The brown dwarf, therefore, moves outward from the hole 
as its mass is tidally stripped. Its effective angular momentum also decreases 
monotonically, so this process is stable on the dynamical time scale.

\subsection{Planets}

Planets like Jupiter \citep{Baraffe, Nettelmann} have complex structures. However, we 
can study two extreme cases, namely, completely solid planets and completely liquid 
ones, in order to see how they react to tidal consumption by a massive hole. 

The densities of solid planets like Earth will remain unchanged when the planets are 
tidally stripped. And these planets tend to have relatively similar densities from the 
center to the surface. Therefore, the mean densities of such planets do not change, 
and the stellar orbital periods do not change either.

Liquid planets (even with a solid core) are self-gravitating spheres following 
polytropic models. For such a planet, its pressure and density roughly follows 
$P = K \rho^{1+1/n}$, where $K$ is a constant, and the polytropic index $n=1$ in this 
case \citep{Kippenhahn}. As the planet is stripped, it still follows the polytropic 
model and satisfies the hydrostatic equations (1) and (2). Therefore we can do simple 
calculations to obtain the new total radius $R_p$ of the planet:
\begin{equation}
   R_p =  \sqrt{\frac{\pi K}{2G}}.
\end{equation}
In other words, the total radius of the planet remains the same all through the stable 
tidal stripping process. So the mean density of the planet decreases, meaning that 
the planet moves out as mass is stripped. Its effective angular momentum decreases as 
well.

\section{Discussion}

The primary motivation for this paper is to understand the response of a star or a 
planet when it is tidally stripped by a massive black hole in a stable manner. Such a 
situation can arise when a star or a planet is formed in an accretion disc, and then 
evolves through circular orbits of diminishing radius under the action of 
gravitational radiation. (Other capture scenarios are possible, and have been widely 
discussed in the literature \citep{Novikov, Sigurdsson, Alexander}.) 

\begin{figure}
 \centering
 \includegraphics[width=3in]{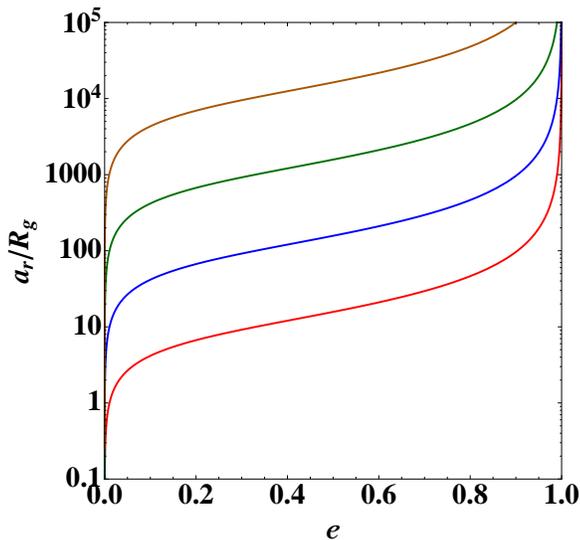}
 \caption{ 
The semi-major axis $a_r$ as a function of the eccentricity $e$ in the decay of a 
two-point mass system, on a log-log scale. Four different colors represent four 
different initial conditions. We set ${a_r}_0$ to be at $10^5 R_g$. Then the initial 
$e_0$'s are (from bottom to top): 0.9999 (red), 0.999 (blue), 0.99 (green), and 0.9 
(orange). }
 \label{eccentricity}
\end{figure}

We constrained the orbit to be circular for simplicity, because of the rapid 
decay of the stellar orbital eccentricity as angular momentum and energy are carried 
away by gravitational radiation. \citet{Peters} gave a detailed calculation of this, 
and showed how the eccentricity changes as the orbit shrinks in the decay of a 
two-point mass system:
\begin{equation}
a_r = \frac{c_0 \ e^{\frac{12}{19}}}{1-e^2} \left( 1+ \frac{121}{304} e^2 
\right)^\frac{870}{2299},
\end{equation}
where $a_r$ is the semi-major axis, $e$ is the eccentricity, and $c_0$ is a constant 
determined by the initial condition $a_r={a_r}_0$ when $e = e_0$. Fig. 
\ref{eccentricity} shows how $e$ varies with $a_r$ for a set of different initial 
conditions.

What is clear from these results is that for most types of stars, the stellar mean 
density decreases as mass is tidally stripped, at least during the later phase of the 
mass-transfer. This, in turn, implies that the orbit will expand after mass transfer 
is initiated. The further implications of this general result will be discussed in a 
forthcoming paper. 

A star in a circular equatorial orbit outside the innermost stable circular orbit 
(ISCO) of the hole will not be swallowed by the hole.  The ISCO of a black hole of 
mass $M$ and spin $a$ has radius $R_{\rm{ISCO}}$ that satisfies the equation:

\begin{equation}
\left(\frac{R_{\rm{ISCO}}}{M}\right)^2-6 \frac{R_{\rm{ISCO}}}{M}+8 a 
\sqrt{\frac{R_{\rm{ISCO}}}{M}}-3a^2=0
\end{equation}
 \citep{Bardeen}. The ISCO resides between $R_g$ (prograde with maximal spin) and 
$9 R_g$ (retrograde with maximal spin).
  
The tidal radius $r_T$ for a star with a mass $M_\star$ and radius $R_\star$ near 
the hole is \citep{Rees}:
\begin{equation}
  r_T =  5 \times 10^{12} \left(\frac{M}{10^6 M_\odot}\right)^\frac{1}{3} 
\frac{R_\star}{R_\odot} \left(\frac{M_\star}{M_\odot}\right)^{-\frac{1}{3}} {\rm cm}\ .
\end{equation}

For a particular black hole, only stars with their tidal radii smaller than the ISCO of 
the hole can get very close to the hole before they are tidally disrupted. For 
example, we can see from Fig. 2 of \citet{Dai} that for a $10^7 M_\odot$ black hole, 
only dwarf stars or lower main sequence stars can get to the ISCO before being tidally 
disrupted. 

Also when the tidal stripping starts, the Roche stellar orbital radius $r_{\rm{R}}$ 
should be larger than the ISCO of the hole. We shall show in an upcoming paper that 
$\bar{\rho}_{\rm{R}} =  M_{\rm{BH}} / (0.683 r_{\rm{R}}^3)$, where 
$\bar{\rho}_{\rm{R}}$ and $r_{\rm{R}} $ are the mean stellar density and orbital 
radius of the star fulfilling the Roche condition. In other words,
\begin{equation}
r_{\rm{R}}  \sim 1.27*10^{13} * \left( \frac{M_{\rm{BH}}}{10^6 M_\odot} 
\right)^{\frac{1}{3}} \left( \frac{\bar{\rho}_\star}
{\bar{\rho}_\odot} \right)^{-\frac{1}{3}} \rm{cm}.
\end{equation}
This radius is just slightly larger than $r_T$ with the same order of magnitude.

\begin{figure}
 \centering
 \includegraphics[width=3in]{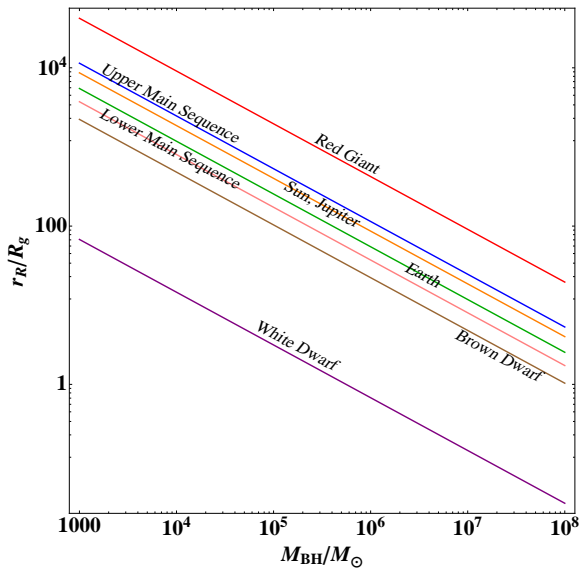}
 \caption{ 
The Roche radius vs central black hole mass for various stars. From top to bottom: a 
red supergiant with a mass $11.85 M_\odot$ (red), an upper main sequence star with a 
mass $\sim 7.9 M_\odot$ (blue), a Sun-like star (orange), a planet like earth (green), 
a lower main sequence star with a mass $\sim 0.3 M_\odot$ (pink), a brown dwarf with a 
mass $0.05 M_\odot$ (brown), and a solar mass white dwarf (purple). A planet like 
Jupiter has a Roche radius similar to that of the Sun-like stars, since their mean 
densities are similar.  }
 \label{RocheRadius}
\end{figure}

Fig. \ref{RocheRadius} shows how the Roche radius varies with the central massive black 
hole mass and the different materials of stars.  We use the set of stars that we 
discuss in this paper. The Roche radius decreases as the star becomes denser, or as 
the hole becomes more massive for most stars or planets. Also this plot shows a 
comparison between a star's Roche radius and the ISCO radius of an SMBH. We can see 
that, for example, the Sun would fill its Roche lobe at a few gravitational radii for a 
$10^7 M_\odot$ SMBH; red giants would always be disrupted before they reach the ISCO 
of SMBHs; and white dwarfs wouldl be tidally disrupted at a few gravitational radii 
for a $10^5 M_\odot$ black hole.

If the mass transfer is initiated when the star is in a relativistic orbit, the 
resulting periodic emission may be detectable by X-ray telescopes, which we will 
discuss in forthcoming papers. Furthermore the behavior of the star could, in principle, 
be monitored as an ``EMRI" source for by a future space-borne gravitational radiation detector. 

\section*{Acknowledgments}

This work was supported by the U.S. Department of Energy contract to SLAC no. DE-AC02-76SF00515. We would like to give special thanks to R. Wagoner, J. Faulkner, and S. Phinney for helpful discussions.


\end{document}